\documentclass[preprint]{revtex4}
\usepackage{natbib}
\usepackage[latin9]{inputenc}
\usepackage{units}
\usepackage{amsmath}
\usepackage{amssymb}
\usepackage{graphicx}
\usepackage{esint}
\usepackage{braket}
\usepackage{multirow}
\usepackage{slashbox}
\usepackage{color}

\makeatletter

\newcommand{\Yb}{\ensuremath{^{171}\mathrm{Yb}^+~}}
\newcommand{\up}{\ensuremath{\left|\uparrow\right\rangle}~}
\newcommand{\down}{\ensuremath{\left|\downarrow\right\rangle}~}
\newcommand{\avg}[1]{\ensuremath{\left\langle#1\right\rangle}}

\makeatother

\begin{document}
\title{Experimental Test of Quantum Jarzynski Equality with a Trapped Ion System}

\author{Shuoming An$^{1}$, Jing-Ning Zhang$^{1}$, Mark Um$^{1}$, Dingshun Lv$^{1}$, Yao Lu$^{1}$, Junhua Zhang$^{1}$, Zhang-qi Yin$^{1}$, H. T. Quan$^{2,3}$, and Kihwan Kim$^{1}$}

\affiliation{$^{1}$Center for Quantum Information, Institute for Interdisciplinary Information Sciences, Tsinghua University, Beijing 100084, P. R. China \\ $^{2}$School of Physics, Peking University, Beijing 100871, P. R. China
\\ $^{3}$Collaborative Innovation Center of Quantum Matter, Beijing, 100871, P. R. China
}

\begin{abstract}
The past two decades witnessed important developments in the field of non-equilibrium statistical mechanics. Among these developments, the Jarzynski equality, being a milestone following the landmark work of Clausius and Kelvin, stands out. The Jarzynski equality relates the free energy difference between two equilibrium states and the work done on the system through far from equilibrium processes. While experimental tests of the equality have been performed in classical regime, the verification of the quantum Jarzynski equality has not yet been fully demonstrated due to experimental challenges. Here, we report an experimental test of the quantum Jarzynski equality with a single \Yb ion trapped in a harmonic potential. We perform projective measurements to obtain phonon distributions of the initial thermal state. Following that we apply the laser induced force on the projected energy eigenstate, and find transition probabilities to final energy eigenstates after the work is done. By varying the speed of applying the force from equilibrium to far-from equilibrium regime, we verified the quantum Jarzynski equality in an isolated system.

\end{abstract}
\date{\today}
\maketitle

There is increasing interest on non-equilibrium dynamics at the microscopic scale crossing over quantum physics, thermodynamics, and information theory as the experimental control and technology at such scale have been rapidly developed. Most of principles in non-equilibrium processes are represented in the form of inequality as seen in the example of the maximum work principle, $\avg{W}-\Delta F \geq 0$, where the average work $\avg{W}$ is equal to the free-energy difference $\Delta F$ only in the case of the equilibrium process. In close to equilibrium processes, the Fluctuation-Dissipation theorem is valid, which connects the average dissipated energy $\avg{W_{\rm diss}}\equiv \avg{W -\Delta F} $ and the fluctuation of the system $\sigma^{2}/2k_{\rm B}T$. Here $\sigma$ is the standard deviation of the work distribution, $T$ is the initial temperature of the system in the thermal equilibrium and $k_{\rm B}$ is Boltzmann constant. Beyond near equilibrium regime, no exact results were known until Jarzynski found a remarkable equality \cite{Jarzynski97} that relates the free-energy difference to the exponential average of the work done on the system:

\begin{eqnarray}
\ln\avg{e^{-W_{\rm diss}/k_{\rm B}T}}=0.
\label{eq:Jarzynski}
\end{eqnarray}

The Jarzynski equality (\ref{eq:Jarzynski}) is satisfied irrespective of the protocols of varying parameters of the system even when the driving is arbitrarily far-from equilibrium. The relation enables us to experimentally determine the $\Delta F$ of a system by repeatedly performing work at any speed. Experimental tests of the classical Jarzynski equality and its related the Crooks fluctuation theorem \cite{Crooks99} have been successfully performed in various systems \cite{Szabo01,Bustamante02,Bustamante05,Rabbiosi05,Ritort05,Blickle06,Harris07,Pekola12,Jarzynski11,Seifert12}.

In classical systems, work can be obtained by measuring the force and the displacement and then integrating the force over the displacement during the driving process. In quantum regime, however, due to the Heisenberg's uncertainty principle, we cannot determine the position and the momentum simultaneously, which invalidates the concepts of the force and the displacement. Instead of measuring these classical observables, it is necessary to carry out projective measurements over the energy eigenstates to determine the work done in each realization and to obtain the work distribution \cite{Hanggi07}. With the understanding of work in quantum mechanics, the Jarzynski equality has been extended to quantum regime \cite{Tasaki00,Kurchan00,Mukamel03} with simplicity and elegancy for the isolated system, though the meaning of work and heat in open quantum systems is still not fully settled \cite{Mukamel09}. 
Although the theoretical derivation of the quantum Jarzynski equality is unequivocal, similar to its classical counterpart, the experimental verification in various systems under various conditions will put it on a solid experimental foundation. For this reason, the experimental testing of the quantum Jarzynski equality has been a long-seeking goal by many physicists  \cite{Huber08,HuberThesis,Talkner11,Serra13,Kehrein12,Vedral13,Paternostro13}. However, even for the isolated quantum system, the experimental verification has been constrained by the technical challenges in controlling quantum systems precisely and performing projective measurements to obtain the work distribution \cite{Huber08,HuberThesis,Talkner11,Serra13}. There have been theoretical efforts to get around such difficulties by measuring the characteristic function and then reconstructing the work distribution \cite{Kehrein12,Vedral13,Paternostro13}. An experimental demonstration has appeared following those proposals \cite{Serra13}. Nevertheless, the standard way of verifying the quantum Jarzynski equality by the direct measurement of the work distribution is still lacking. Here, we adopt the methods of two measurements over energy eigenstates and obtain the work distribution. From the work distribution, we verify the quantum Jarzynski equality.

In our experiment, we employ a trapped atomic \Yb ion harmonic oscillator whose Hilbert space has infinite dimensions. We implement the projective measurement on phonons \cite{CShen14} to determine the initial eigenstate from the thermal distribution and perform the standard phonon-distribution measurement \cite{Meekhof96,DidiRMP,Walther12,Bowler12} after work is done on the projected eigenstate to obtain the work distribution. We test the quantum Jarzynski equality at various initial temperatures and switch speeds. We compare the performance of the Jazynski estimate of $\Delta F$ to that by the average work and the Fluctuation-Dissipation theorem.

\begin{figure}[ht]
  \includegraphics[width=1\columnwidth]{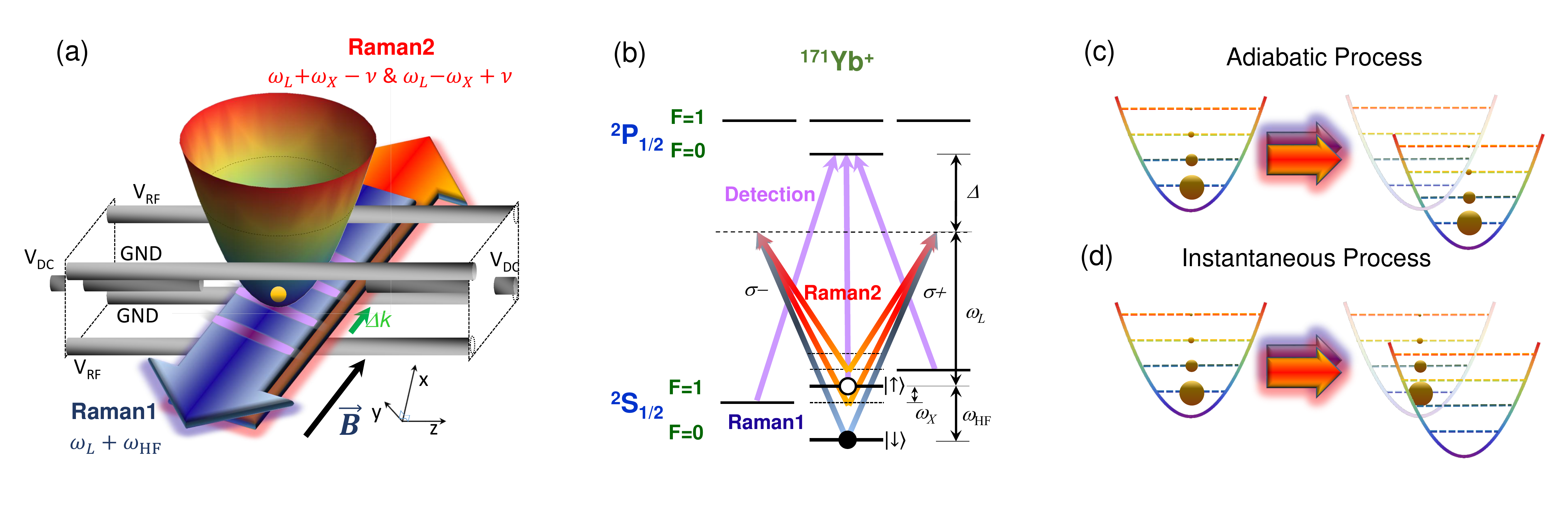}\\
  \caption{Experimental setup for the test of the Jarzynski equality and equilibrium and non-equilibrium work processes. (a) Schematic of the ion trap apparatus as an ideal harmonic oscillator and the geometry of laser beams that generate an effective moving standing wave which pushes the ion. The counter propagating laser beams drive transitions between states in \Yb ion shown in (b). (b) The basic level structure of \Yb ion and the relevant laser frequencies. The Raman laser beams introduce the state-dependent force. When the beat-note frequencies of them are adjusted to near $\omega_{\rm HF}\pm \omega_{X}$, it pushes the ion along $\pm\Delta k$ direction for the $(\ket{\uparrow}\pm\ket{\downarrow})/\sqrt{2}$ state of the ion. (c) The phonon distribution before and after the perfect adiabatic process dose not change. (d) For the non-equilibrium process, the final phonon states are widely distributed. In both cases of the adiabatic and the instantaneous shifts of the harmonic oscillator, the Jarzynski equality should be valid, as long as the system is initially prepared in a thermal equilibrium state. }\label{fig1:setup}
\end{figure}

The ion trap system has shown a high degree of control in quantum regime \cite{DidiRMP}. Controls of the harmonic oscillator are performed through the coupling to the two electronic levels (qubit) of \Yb ion in the $S_{1/2}$ manifold denoted by: $\ket{F=1, m_F=0}\equiv\up$ and $ \ket{F=0, m_F=0} \equiv \down$that are separated by $\omega_{\rm HF} = (2\pi) 12.642$ GHz. As shown in Fig. \ref{fig1:setup}, the \Yb ion is confined in harmonic potentials with the trap frequencies $\omega_{X}$=$(2\pi)3.1$ MHz, $\omega_{y}$=$(2\pi)2.7$ MHz, $\omega_{z}$=$(2\pi)0.6$ MHz, respectively.

We perform work on the system by applying a laser induced force and shifting the center of the potential in the $X$-direction. The force is implemented by counter-propagating laser beams shown in Fig. \ref{fig1:setup}(a) and (b), which is equivalent to generating the \emph{so-called} qubit-state dependent force \cite{Haljan1ion,PLee05}. The pair of laser beams with the frequency differences of $\omega_{\pm}=\omega_{\rm HF}\pm(\omega_{X}-\nu)$ produce the following Hamiltonian in the rotating frame about $H_{0}=\frac{1}{2}\hbar \omega_{\rm HF} \hat{\sigma}_{z}+ \hbar (\omega_{X}-\nu)\left(\hat{a}^{\dagger} \hat{a}+\frac{1}{2}\right) $ after taking the rotating-wave approximation,
\begin{eqnarray}
H\left(t\right)=\frac{\hat{P}^2}{2 M_{e}}+\frac{1}{2} M_{e} \nu^2 \hat{X}^2 + f(t) \hat{X} \hat{\sigma}_{x}.\label{eq:Ham}
\end{eqnarray}
Here $M_{e}= \frac{\omega_{X}}{\nu} M$ is the scaled mass of the \Yb mass, $M$, $\nu [\equiv \omega_{X} \pm (\omega_{\rm HF}-\omega_{\pm})]  =(2\pi) 20.0$ kHz is the effective trap frequency, $f(t)=\frac{1}{2}\hbar \Delta k \Omega (t)$ is the effective force and $\hat{\sigma}_{x}$ is the Pauli operator. The $\Delta k$ is the net wave-vector of the counter-propagating laser beams along the $X$-axis and the Rabi frequency $\Omega$ is proportional to the intensity of the laser beams.

The force shifts the trap center by $-f(t)/M_{e}\nu^2$ and reduces the ground state energy by $f^2(t)/2M_{e}\nu^2$. In our experiment, the maximum force is 4.16 zN ($\times 10^{-21}$N) produced by the maximum Rabi frequency $\Omega_{\rm max}= (2\pi) 378$ kHz, which shifts the center position by 5.6 nm. When we adiabatically add the force $f(t)$ to the maximum value, the final state distribution is conserved in the new basis and still in the thermal equilibrium shown Fig. \ref{fig1:setup}(c). On the contrary, if we increase the force to the same value instantaneously, the final states are highly excited, which represents a far-from equilibrium process and is shown in Fig. \ref{fig1:setup}(d). In both cases, we would measure the same average of the exponentiated dissipated work $\avg{\exp{(-W_{\rm diss}/k_{\rm B}T)}}$, which is used to test the Jarzynski equality.

For the time-dependent quantum system $H(t)$ (\ref{eq:Ham}), where the eigenvalues and the eigenstates are denoted by $E_{n}(t)$ and $\ket{n(t)}$, the phonon number state, the work done on the system from $t=0$ to $t=\tau$ is defined by $E_{\bar{n}}(\tau)-E_{n}(0)$. The distribution of the work is described by the following equation \cite{Hanggi07}
\begin{eqnarray}
P\left(W\right)=\sum_{n,\bar{n}} \delta[W-\left(E_{\bar{n}}(\tau)-E_{n}(0)\right)]P_{\bar{n}\leftarrow n}P^{th}_{n},
\label{eq:PW}
\end{eqnarray}
where $P^{th}_{n} = \exp{\left(-E_{n}(0)/k_{\rm B}T\right)}/\left[\sum_{n}{\exp{\left(-E_{n}(0)/k_{\rm B}T\right)}}\right]$ shows the initial thermal distribution and $P_{\bar{n}\leftarrow n}$= $|\bra{\bar{n}(\tau)}\hat{U}\ket{n(0)}|^2$ is the transition probability from the initial state $\ket{n(0)}$ to the final state $\ket{\bar{n}(\tau)}$ under the evolution operator $\hat{U}$. For the test of the validity of the Jarzynski equality, it is necessary to observe that the average of the exponentiated work $\avg{\exp{(-W/k_{\rm B}T)}}\equiv\sum{P\left(W\right) \exp{(-W/k_{\rm B}T)}}$ does not depend on the protocol of applying the work from quasi-static to far-from equilibrium regime. The essential part of the experimental test in quantum regime is to obtain the conditional probability from the projected energy eigenstate $\ket{n(0)}$ out of the thermal distribution to the final eigenstate $\ket{\bar{n}(\tau)}$ after the work is done on the projected state.

In our experiment, we follow a similar procedure to that proposed in Ref. \cite{Huber08, HuberThesis,Talkner11}, which is composed of four stages: (i) preparation of the thermal state; (ii) projection to an energy eigenstate; (iii) application of work on the eigenstate; (iv) the measurement of the final phonon distribution.

\begin{figure}[ht]
  \includegraphics[width=0.5\columnwidth]{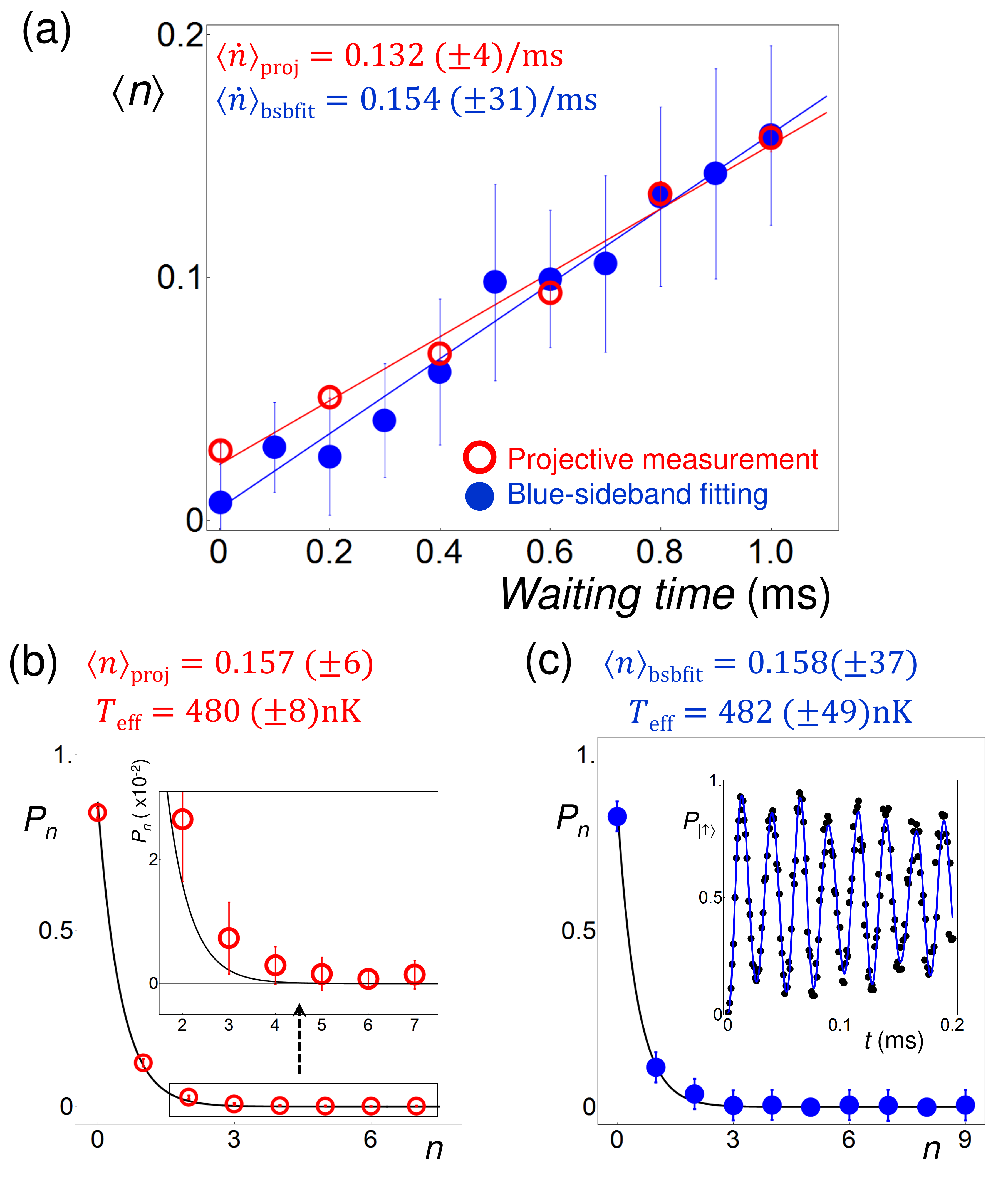}\\
  \caption{Thermal state preparation of the phonon and the measurement results. (a) Measurements on the heating rate of the trapped ion's harmonic motion in the $X$-direction. After the ground state cooling, the average phonon number increases linearly with the waiting time. The temperature is extracted from the average phonon number $\avg{n}$ by $T_{\rm eff}=\hbar \nu /k_{\rm B} \ln{\left(1+1/\avg{n}\right)}$, where $\nu$ is the effective trap frequency. The projective measurements on the phonon states (red circles) are performed at various waiting times and the results are compared to those from the phonon-distribution measurement based on the dependence of the blue-sideband Rabi frequency on the phonon number shown in Eq. (\ref{eq:Pup}) (filled blue circles). The measured heating rate by the phonon projection is consistent with that by fitting the blue-sideband transition by Eq. (\ref{eq:Pup}) within error bars. (b) The phonon distribution (red circles) measured by projective measurement at 1 ms waiting time, fitted by the thermal distribution function $P_{n}^{th} = \avg{n}^n/(\avg{n}+1)^{n+1}$ (solid line). The measurement procedure shown in Fig. \ref{fig3:Projection} is repeated by $5\times 10^{6}$ times, where the phonon state is determined at each single run in the projective measurement. Here, the experimentally determined errors by the detection are corrected (see Methods). The error bar is the standard deviation. (c) The phonon distribution at 1 ms obtained by fitting the data in the inset with a superposition of Fock states (filled blue circles) and by a thermal distribution with $\avg{n}$ (solid line) through Eq. (\ref{eq:Pup}). The error bars come from the one standard deviation in parameter estimations of the fitting. The method is applied to obtaining the error bars for the rest of figures and table (for more detail, see the supplementary information). }\label{fig2:Thermal}
\end{figure}

We prepare a thermal state of the trapped ion's harmonic motion in the $X$-direction. We first cool the vibrational mode near to the ground state $\ket{n=0}$ by the resolved sideband cooling \cite{Monroe95a} right after the Doppler cooling. Then we let the system be heated up to the desired temperature. The heating and the thermalization of the system have been extensively studied both experimentally and theoretically \cite{Turchette00,Wineland00,Myatt00,Messina03}. The process is well described by the model of a harmonic oscillator coupled to a high temperature reservoir, which manifests the thermal distribution at any instant of the heating process \cite{Turchette00,Wineland00,Myatt00,Messina03}. We observe the thermal distribution at each waiting time, where the temperature increases as shown in Fig. \ref{fig2:Thermal}(a). We characterize the distribution by both the projective measurement and the standard fitting method using Rabi-frequency dependence of the sideband transition on the phonon state \cite{Meekhof96,DidiRMP,Walther12,Bowler12}.

Fig. \ref{fig2:Thermal}(b) shows the phonon distribution of a thermal state obtained by the projective measurement at 1 ms waiting time after the ground state cooling, which  yields $\avg{n}=0.157 (\pm 6)$, $T_{\rm eff}=480 (\pm 8)$ nK. The temperature is extracted from the average phonon number $\avg{n}$ by $T_{\rm eff}=\hbar \nu /[k_{\rm B} \ln{\left(1+1/\avg{n}\right)}]$, where the effective frequency of the harmonic oscillator $\nu$ is different from the real motional frequency $\omega_{X}$. We use the thermal state at 1 ms as well as 0.3 ms ($\avg{n}=0.051  (\pm 2)$, $T_{\rm eff}=316 (\pm 4)$ nK) and 0.6 ms ($\avg{n}=0.094  (\pm 4)$, $T_{\rm eff}=390 (\pm 6)$ nK) to start with different initial temperatures for the test of the Jarzynski equality. We finally determine the population of each eigenstate by fitting with the thermal distribution due to the limited precision of the projective measurement below $10^{-3}$ level (see Method and supplementary information). The result of phonon distribution by the projective measurement is consistent with that from the fitting method as shown in Fig. \ref{fig2:Thermal}(c). Note that in the fitting method, we cannot decide the phonon state in a single measurement and cannot carry out the successive operations on the projected Fock state.

\begin{figure}[ht]
  \includegraphics[width=1\columnwidth]{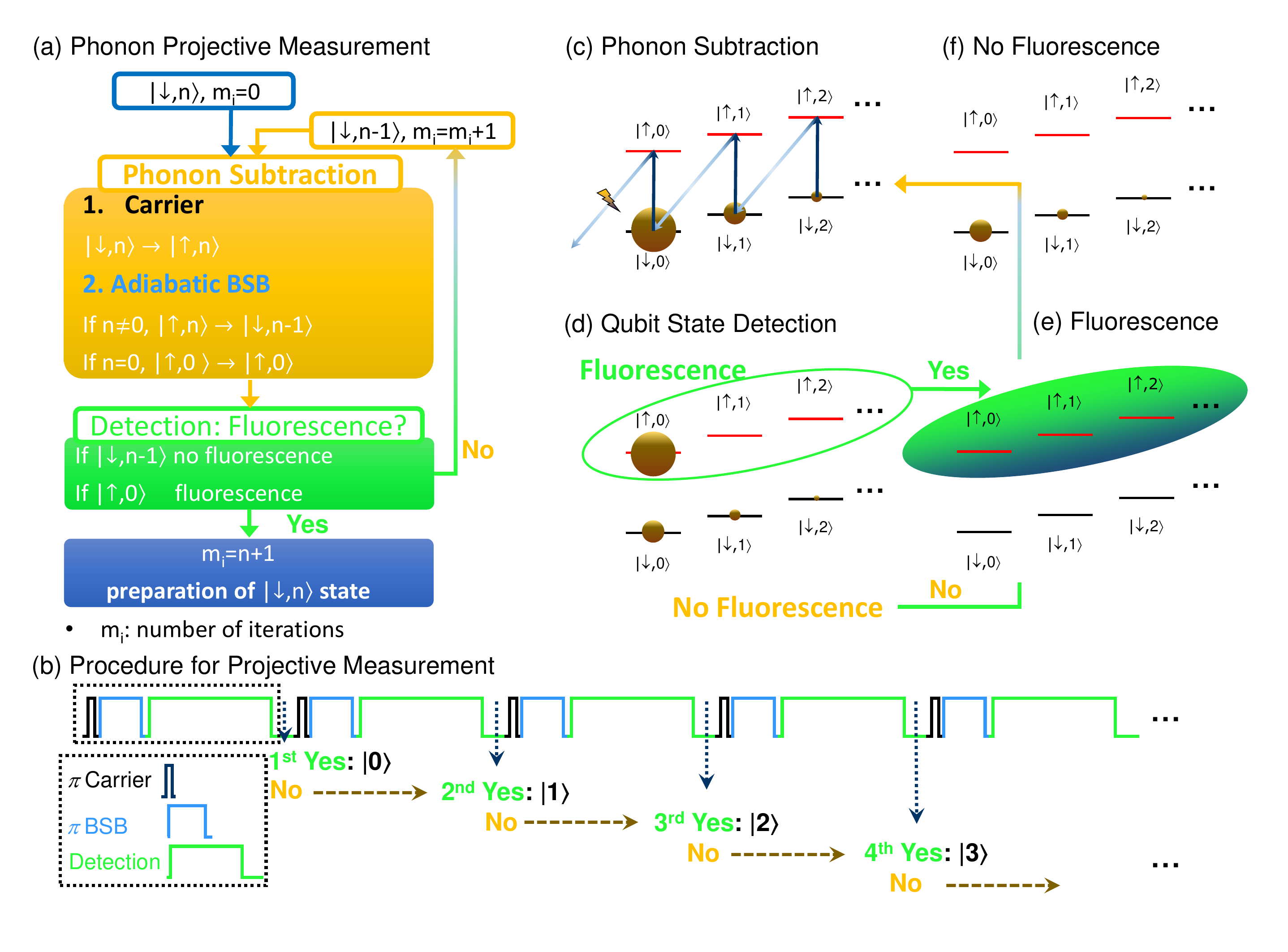}\\
  \caption{Experimental scheme for the projective measurement of the phonon state. (a) The flow chart and (b) the time sequence for the projective measurement. If the first fluorescence is observed after $(n+1)$ times repetitions of (c) the phonon subtraction and (d) the qubit-state detection, the projected phonon state is $\ket{n}$. (c) The phonon subtraction operation that changes the phonon state from $\ket{n}$ to $\ket{n-1}$, is composed of the $\pi$ pulses of the resonant carrier transition (black arrow) and the adiabatic blue-sideband (BSB) transition (blue arrow, see also Method section). The subtraction $\ket{n}\rightarrow\ket{n-1}$ is performed for any phonon Fock state except $\ket{n=0}$, which is transferred to $\ket{\uparrow, n=0}$. (d) On the application of the detection laser beam [Fig. \ref{fig1:setup}(c)], (f) fluorescence is observed on \up state, which was $\ket{\downarrow, n=0}$ state before the phonon subtraction and (g) no fluorescence is detected on \down state, where phonon states are reduced by one quanta. The state originally projected to $\ket{\downarrow, n}$ reaches to the state $\ket{\uparrow, n=0}$ that shows fluorescence after ($n+1$)-times repetitions of (c)(d) procedures.}\label{fig3:Projection}
\end{figure}

Our projective measurement is composed of two parts: firstly, find the projected energy eigenstate $\ket{n}$ and then deterministically generate the same phonon Fock state. The projected state is determined by repeating the sequence of the phonon subtraction and the qubit-state detection and counting the number of repetitions when the first fluorescence is observed [Fig. \ref{fig3:Projection}(a)]. If the fluorescence is detected at the $(n +1)$-th iteration for the first time, the projected state is $\ket{n}$ [Fig. \ref{fig3:Projection}(b)] \cite{CShen14}.

The phonon subtraction ($\ket{n}\rightarrow\ket{n-1}$, shown in Fig. \ref{fig3:Projection}(c)) is performed by the successive application of the $\pi$ pulse of the resonant carrier transition ($\ket{\downarrow, n}\rightarrow\ket{\uparrow,n}$) and the adiabatic blue-sideband transition ($\ket{\uparrow, n}\rightarrow\ket{\downarrow,n-1}$). We apply the scheme of the adiabatic spin-flip operation in Ref. \cite{Junhua14} to the phonon system, where the near-perfect transition from $\ket{\uparrow, n}$ to $\ket{\downarrow,n-1}$ with the same duration is achieved regardless of the phonon number $n$ (see Method section and supplementary information). After the phonon subtraction, only the state $\ket{\downarrow,0}$ is transferred to the bright \up state that scatters photons at the qubit-detection sequence [Fig. \ref{fig3:Projection}(e)] and the other states remain in the dark \down state with one motional quanta reduced [Fig. \ref{fig3:Projection}(f)]. Therefore, the projected state $\ket{\downarrow,n}$ generates the fluorescence at the $(n+1)$-th successive operation of the subtraction and the detection. At each waiting time, we typically perform $5\times 10^{6}$ projective measurements with 7 iterations of the subtraction and the detection. At each measurement, the projected energy eigenstate is determined. The Fock state $\ket{n}$ is deterministically prepared by the $n$ times application of the $\pi$ pulses of the blue-sideband and the carrier transition after another ground state cooling \cite{Meekhof96}.

\begin{figure}[ht]
  \includegraphics[width=0.7\columnwidth]{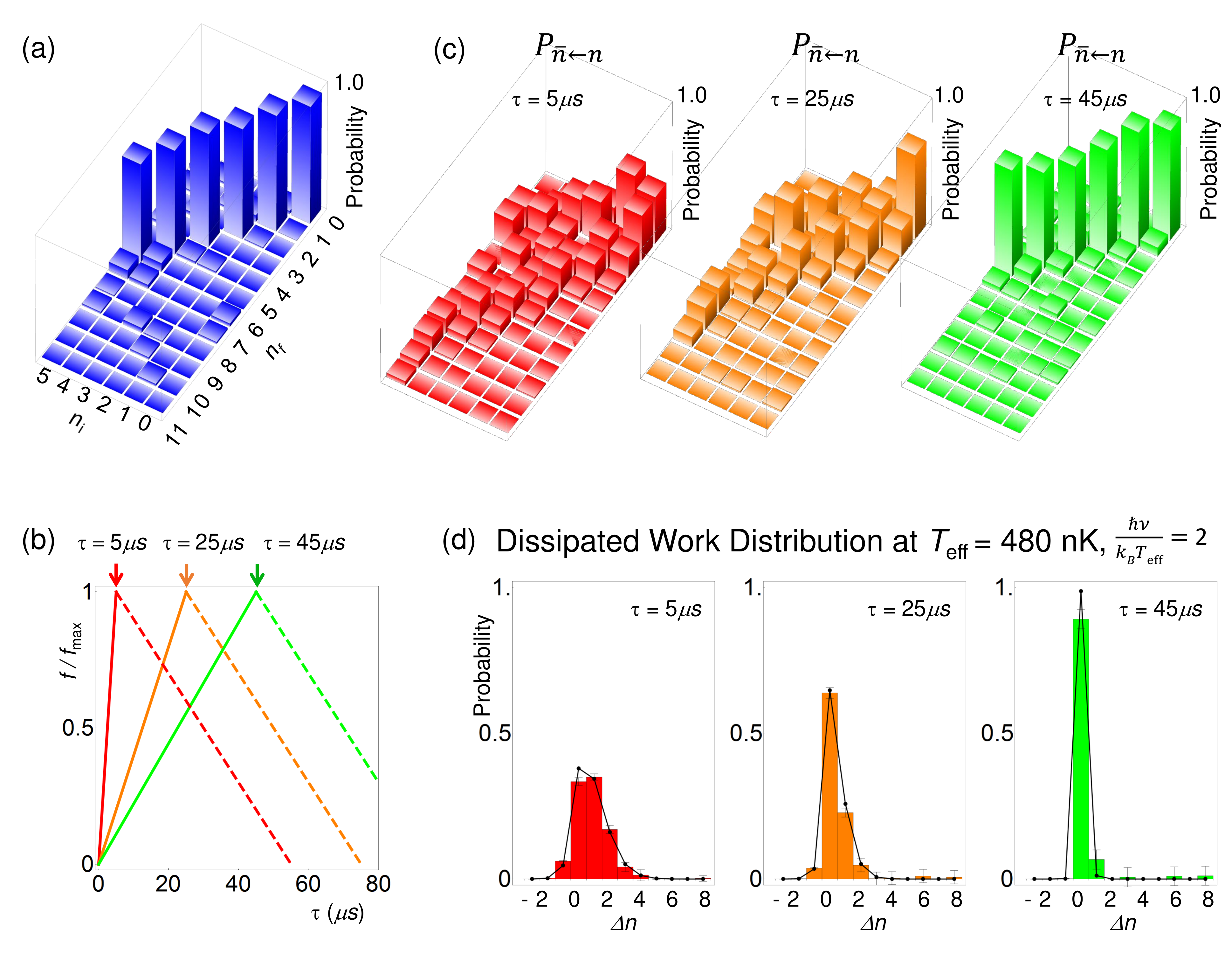}\\
  \caption{The dissipated works $W_{\rm diss}=W-\Delta F=\hbar \nu \Delta n$ and the probabilities $P_{\bar{n}\leftarrow n}$ of three different speeds of ramping up the force. (a) The populations of the prepared phonon Fock states from $n=0$ to 5 (see also supplementary materials). (b) The force is linearly increased to the maximum value in three different durations, which are corresponding to the far-from equilibrium (5 $\mu$s), the intermediate (25 $\mu$s) and the near adiabatic (45 $\mu$s) processes of work. The dashed line shows the adiabatic process to bring the system to the lab frame in $50~\mu$s. (c) The transfer probabilities from $\ket{n(0)} (n(0)=0,1,\cdots,5)$ to $\ket{\bar{n}(\tau)}$ with three speeds measured by the maximal-likelihood method of the blue and the red sideband transitions after the work process (see also supplementary information). The time evolutions by the blue and the red transitions are monitored up to 250 $\mu$s with 1 $\mu$s interval by averaging 200 repetitions at each step. (d) The distributions of the dissipated works at $\avg{n} =0.157$ have full information to test the validity of the quantum Jarzynski equality, Eq. (\ref{eq:Jarzynski}). The data (bars) are resulted from the transfer probability (c) with the weight of thermal distributions by projective measurements shown in Fig. \ref{fig2:Thermal}(b). The solid lines are from the analytical calculations. The three distributions of dissipated works show the characteristics of far-from equilibrium, the intermediate and the adiabatic processes, as the width increases when the process is changed to the far-from equilibrium. The work distribution for the $5 \mu$s duration clearly shows a non-Gaussian profile, which has quantum origin. For the case of 45 $\mu$s driving, the distribution is close to a delta function.}\label{fig4:Results}
\end{figure}

Since the projection scheme has two totally independent sequences: determination and preparation, we completely separate the sequence of preparing the Fock state $\ket{n}$ from the sequence of detecting it. We prepare the initial Fock state up to $\ket{n=5}$, since the total population of larger than $n=5$ Fock states at $\avg{n}=0.157$ is less than $10^{-5}$ and the estimated error without these states is smaller than $10^{-3}$ even for the case of the fastest driving. Fig. \ref{fig4:Results}(a) shows the fidelity in the preparation of the phonon number state up to $\ket{n=5}$. We produce the $\ket{n}$ state up to $n=5$ with over 90$\%$ fidelity (see supplementary information).

After determining and preparing the projected energy eigenstate $\ket{n}$, we provide work on the system. We apply the laser induced force of Eq. (\ref{eq:Ham}) on the prepared state for the durations of 5 $\mu$s, 25 $\mu$s and 45 $\mu$s with the linear increase of the strength to the same maximum value as shown in Fig. \ref{fig4:Results}(b). The qubit-state dependence in the Hamiltonian (\ref{eq:Ham}) does not play any role, since the electronic state is prepared in the eigenstate of $\hat{\sigma}_{x}$, $(\ket{\uparrow}\pm\ket{\downarrow})/\sqrt{2}$  (see Method section). We have to note that the laser induced effective force occurs in the rotating frame. Since the effective force term does not commute with $H_{0}$, we adiabatically bring the final state to the lab frame before the measurement of the final phonon distribution. We've carefully studied the adiabatic process and observed less than 0.015 change of the average phonon number when the total duration is 50 $\mu$s including the heating effect (see Methods section and the supplementary information).

After the work is done on the state $\ket{n}$, we measure the final distribution of phonon number states by applying the blue and the red sideband transitions and fitting the signals through the maximal likelihood method with the parameters of the Fock state population $P_{n}$ (see Method section and supplementary information). We observe the time evolution of the blue and the red sideband transitions up to 250 $\mu$s with 1 $\mu$s step by averaging 200 repetitions of each step.

Fig. \ref{fig4:Results}(c) summarizes the final phonon distributions depending on the speed of applying work on a Fock state $\ket{n}$, which are the transition probabilities $P_{\bar{n}\leftarrow n}$ in Eq. (\ref{eq:PW}). The raw data and the fitting results are presented in the supplementary information. Fig. \ref{fig4:Results}(c) shows that our protocols of work range from equilibrium to non-equilibrium regime. For the slow ramp, $\tau = 45~\mu$s, the final distribution of the phonon states is almost identical to the initial distribution shown in Fig. \ref{fig4:Results}(a). For the case of the fastest ramping, $\tau = 5~\mu$s, it is clearly shown that the final populations spread the most widely, which indicates the process is out of equilibrium.

Fig. \ref{fig4:Results}(d) shows the probability distribution of the dissipated work, $W_{\rm diss}$, constructed from the transfer probabilities $P_{\bar{n}\leftarrow n}$ shown in Fig. \ref{fig4:Results}(c) with the phonon distribution $P^{th}_{n}$ of effective temperature $T_{\rm eff}=480$ nK ($\avg{n}$=0.157). It is clear that the ramping of the force with the duration $\tau$ = 45 $\mu$s is close to the adiabatic process, since there is almost no change of the phonon distribution in the work process. Similar to the results of classical regime \cite{Bustamante02}, the mean value and the width of the distribution of the dissipated work increase with the pulling speed. Note that in the classical version of our model, only Gaussian profile of the distribution of dissipated work exists in an open system \cite{Jarzynski99,Speck04,Sasso07} as well as in an isolated system (see also supplementary information) regardless of the protocol. The non-Gaussian profile is predicted in Ref. \cite{Talkner08} at the extreme low initial temperature $\avg{n}<1$, which is naturally in the quantum regime.

Table \ref{Tab:Results} summarizes the performances of the three different estimates of the free energy changes from the Jarzynski equality, the Fluctuation-Dissipation theorem, and the mean dissipated energy. Our experimental data clearly demonstrate the validity of the Jarzynski equality when other estimates deviate from the ideal values \cite{Jarzynski01}. For comparison, we apply three work protocols at three initial temperatures. The free energy differences $\Delta F$ are -2.63 $k_B T_{\rm eff} ~(T_{\rm eff}=316 \rm nK)$, -2.13 $k_B T_{\rm eff} ~(T_{\rm eff}=390 \rm nK)$, and -1.73 $k_B T_{\rm eff} ~(T_{\rm eff}=480 \rm nK)$. For the case of $\tau=45 \mu$s, the average dissipated energy provides the estimations of $\Delta F$ within 2$\sigma$. For the case of $\tau=45 \mu$s and $\tau=25 \mu$s, the Fluctuation-Dissipation theorem provides the estimations within $\sigma$ and 2$\sigma$, respectively, which indicates these protocols are in linear response regime. For the case of $\tau=5 \mu$s  drive, the far-from equilibrium protocol, only the Jarzyski estimate gives the reasonable values of the free-energy differences. We found that the main error in the experiment comes from the heating of the phonon mode during the adiabatic return, but the amount of error from the heating in the Jarzynski estimate is less than the experimental uncertainty according to our numerical simulation. The detailed experimental imperfections and errors are discussed in the supplementary information.

For many decades, the measurement of the work and its distribution in a quantum system has been only a thought experiment and this fact may explain why the attempts of testing the quantum Jarzynski equality in experiments has not been successful so far. Based on the ground-breaking technology of isolating and manipulating individual quantum systems \cite{Wineland13,Haroche13}, developed in the last three decades, we further developed the phonon projective measurement method, which enables us to experimentally measure the mechanical work and its distribution in a quantum system undergoing an arbitrary non-equilibrium process. Besides being used for verifying the quantum Jarzynski equality, our experimental breakthrough could be applied to make many other thought experiments in quantum thermodynamics tangible in the laboratory. For example, the method can be immediately used to the test of Crooks' relation in quantum regime \cite{Crooks99,Bustamante05} and to further investigations of the equality in an open quantum system \cite{Mukamel09,Talkner11}. It could also be adapted to the studies of quantum heat engines \cite{Quan07,Lutz12} by experimentally exploring work and heat in thermodynamic cycles. In addition, the phonon projective measurement would be the essential tool for the Boson-sampling problem \cite{Aaronson11} with phonons \cite{CShen14}.

\section*{Methods}
In the projected measurement, we transfer the $\ket{\uparrow, n}$ state to the $\ket{\downarrow, n-1}$ state in the same duration independent of the phonon number by adjusting the intensity and the frequency of the laser beam with the shape of $\Omega_{b}(t)=\Omega_{n,n+1, \rm max}\sin(\frac{\pi}{T}t)$ and $\delta(t)=\delta_{0}\cos(\frac{\pi}{T}t)$\cite{Junhua14}. The main errors in the phonon projective measurement come from two sources: the imperfection in the qubit-state detection and the heating of the phonon state during the measurement. We apply the correction method of the state detection \cite{Chao12} to the phonon system. The phonon state changes due to the heating process are tracked and reversed by a calculation. We apply the $\hat{\sigma}_{x}$-dependent force and we prepare the eigenstate of $\hat{\sigma}_{x}$ during the application of the force and the adiabatic process back to the lab frame. We bring the phonon distribution from the rotating frame to the lab frame by linearly reducing the laser intensity in $T_{a}=\frac{2\pi}{|\nu|}$. We utilize the maximal-likelihood method to find the phonon distributions by fitting the interferences of the blue and the red sideband transitions among different phonon modes following the methods in \cite{PoschingerThesis}. The $P_{n}$ can be determined by observing the time evolution of $P_{\uparrow}^{\pm} (t)$ of finding the ion in the $\up$ state during the blue (red) sideband transitions, which is written as
\begin{eqnarray}
P_{\uparrow}^{\pm} \left(t\right) =\frac{1}{2} \sum_{n} P_{n} \left[1- e^{-\gamma_{\pm} t}A_{\pm}\cos\left(2 \Omega_{n,n\pm 1} t\right)\right]. \nonumber
\label{eq:Pup}
\end{eqnarray}
Here $P_{\uparrow}^{\pm} (t)$ is experimentally measured through the qubit-state dependent fluorescence \cite{Olmschenk07,Zhang12}. The decay rate $\gamma_{\pm}$ is included due to the laser intensity fluctuation and the heating. The contrast $A_{\pm}$ is required to take into account of the imperfection in the state preparation and detection. The Supplementary Information describes the detailed experimental schemes with supporting data.

\subsection*{Acknowledgment}
This work was supported in part by the National Basic Research Program of China Grant 2011CBA00300, 2011CBA00301, the National Natural Science Foundation of China Grant 61073174, 61033001, 61061130540, 11374178, 11375012 and 11105136. KK and HTQ acknowledge the recruitment program of global youth experts of China.


Correspondence should be addressed to H.T.Q (htquan@pku.edu.cn) and K.K (kimkihwan@mail.tsinghua.edu.cn).

\newpage

\begin{table}
\begin{centering}
\caption{The summary for the experimental test of the quantum Jarzynski equality and the comparison to other estimates. All the experimental results at various temperatures and the speeds of force application are close to the ideal values of $-\ln\avg{e^{-W_{\rm diss}/k_{\rm B}T}}=0$ within error bars, where $W_{\rm diss}=W-\Delta F$ and the energy scale is $k_{\rm B} T_{\rm eff}$. The other estimates of the free-energy differences $\Delta F$ by the average work $\avg{W_{\rm diss}/k_{\rm B}T_{\rm eff}}$ and by the Fluctuation-Dissipation relation $\avg{W_{\rm diss}/k_{\rm B}T_{\rm eff}}-\frac{1}{2}\frac{\sigma^2}{(k_{\rm B} T_{\rm eff})^2}$ are compared to those by the Jarzynski equality \cite{Jarzynski01}. For these estimates from experimental results, the small populations of high phonon states significantly smaller than error bars are not included.}
\label{Tab:Results}
\scalebox{0.7}{
\begin{tabular}{c|c|c|c||c|c|c||c|c|c}
\noalign{\hrule height 0.8pt}
				$\Delta F/k_{\rm B} T_{\rm eff}$ & \multicolumn{3}{c||}{$-\ln\avg{e^{-W_{\rm diss}/k_{\rm B}T_{\rm eff}}}$}& \multicolumn{3}{c||}{$\avg{W_{\rm diss}/k_{\rm B}T_{\rm eff}}-\frac{1}{2}\frac{\sigma^2}{(k_{\rm B} T_{\rm eff})^2}$}& \multicolumn{3}{c}{$\avg{W_{\rm diss}/k_{\rm B}T_{\rm eff}}$}\\
\cline{2-10}
									& $\tau=$5 $\mu$s &  $\tau=$25 $\mu$s  & $\tau=$45 $\mu$s & $\tau=$5 $\mu$s &  $\tau=$25 $\mu$s  & $\tau=$45 $\mu$s & $\tau=$5 $\mu$s &  $\tau=$25 $\mu$s  & $\tau=$45 $\mu$s \\
\noalign{\hrule height 0.8pt}
-2.63 (316 nK)  &  -0.032($\pm$37) & 0.006($\pm$34)	& 0.042($\pm$52) &  -1.601($\pm$443) & -0.718($\pm$568)	& -0.087($\pm$154) &  2.573($\pm$313) & 0.929($\pm$401)	& 0.211($\pm$109) \\
-2.13 (390 nK) &  -0.033($\pm$35) & 0.005($\pm$33)	& 0.037($\pm$50) &  -0.889($\pm$346) & -0.426($\pm$442)	& -0.027($\pm$120) &  2.033($\pm$245) & 0.749($\pm$313)	& 0.168($\pm$85)\\
-1.73 (480 nK) &  -0.034($\pm$34) & 0.003($\pm$31)	& 0.031($\pm$48) & -0.505($\pm$269) & -0.260($\pm$342)	& 0.002($\pm$93) &  1.598($\pm$190) & 0.602($\pm$242)	& 0.131($\pm$66) \\
\noalign{\hrule height 0.8pt}
\end{tabular}
}
\par\end{centering}
\end{table}

\end{document}